\documentclass{aa}
\usepackage{epsfig}
\usepackage{rotating}

\usepackage{graphicx}

\newcommand{\rrls}{RR\,Lyrs}
\newcommand{\rrl}{RR\,Lyr}
\newcommand{\ea}{et~al.~}
\newcommand{\hip}{{\sc Hip}}

\begin{document}
   \title{Proper identification of RR Lyrae Stars brighter than 12.5 mag}

   \author{G. Maintz 
 }

\offprints{Gisela Maintz, 
\email{gmaintz@astro.uni-bonn.de}}

\institute{Sternwarte der Universit\"at Bonn, Auf dem H\"ugel 71,
              D-53121 Bonn, Germany \\
}

\date{Received .../ Accepted ...}
  {\date{Received 12.4.2005; Accepted 13.7.2005}}

\abstract{ RR Lyrae stars are of great importance for investigations 
of Galactic structure. 
However, a complete compendium of all RR-Lyraes in the solar neighbourhood 
with accurate classifications and coordinates does not exist to this day. 
Here we present a catalogue of 561 local RR-Lyrae stars
($V_{\rm max}  \leq 12.5$ mag) according to the magnitudes given
in the Combined General Catalogue of
Variable Stars (GCVS) 
and 16 fainter ones.
The Tycho2 catalogue 
contains $\simeq 100$ RR Lyr stars. 
However, many objects  have 
inaccurate coordinates in the GCVS, 
the primary source of variable star information, 
so that a reliable cross-identification is difficult.
We identified RR Lyrae from both catalogues based on 
an intensive literature search. 
In dubious cases we carried out photometry of fields to identify the variable. 
Mennessier \& Colome (2002) have published a paper with 
Tyc2-GCVS identifications, 
but we found that many of their identifications are wrong. 
\keywords{astrometry --  Stars:  RR Lyrae stars -- Catalogues: 
Tycho-2 catalogue  
--  Catalogues: The HST Guide Star Catalogue, Version 1.2 
-- Catalogues:   Combined General Catalogue of Variable Stars}
}

\maketitle
\section{Introduction}
\label{intro}

 RR Lyrae stars (\rrls) are classical variables of the instability strip. 
They are one of the primary distance indicators, 
which is why they have often been investigated 
to determine their absolute magnitude and their distances 
(see, e.g., references in Fernley \ea 1998). 
In addition, \rrls\ provide  an excellent sample of stars to investigate 
kinematics  in the galaxy (see, e.g., Beers \ea 2000, 
Layden 1994, Layden \ea 1996, 
Dambis \& Rastorguev 2001, Martin \& Morrison 1998, Altmann \& de Boer 2000).

Several studies have recently been carried out 
on other horizontal-branch (HB) stars. 
Altmann \ea (2004) published a study of the orbits and halo distribution 
of sdB stars, extending earlier work by de Boer \ea (1997).
A similar study on RHB stars has been completed, too (Kaempf \ea 2005).
It thus appeared worthwhile to also investigate the orbits 
and the ensuing $z$-distribution of \rrls\ (Maintz \& de Boer 2005).

For such purposes the distance of the stars as well as 
accurate data of proper motion and radial velocity are required.
A good source for astrometric data is the  
Tycho-2 catalogue (H{\o}g \ea 2000; Tyc2) 
since it provides astrometric data for more than 2.5 million stars.
However, one first has to identify the \rrls\ in that catalogue. 

We took the Combined General Catalogue of Variable Stars 
(Kholopov \ea 1998, GCVS) as a basic source for compiling an unbiased sample 
of \rrls. The GCVS is, as the authors write,
``the only reference source on all known variable stars...
The total  number of designated variable stars has now reached 31918.''
However, 
the coordinates given in GCVS are not as accurate as one would like for, 
specifically, cross-correlations with other catalogues.
The positions in GCVS are said to be accurate ``to 1 second of arc''.
But we found that this accuracy is too optimistic.
In some cases the coordinates denoted not the variable but another star. 
In even more cases, at the location given by the coordinates 
there was no star at all.
According to the authors of GCVS,
 ``the Sternberg Institute has started preparation of an electronic
    release GCVS 4.2 which will contain improved light elements,
    classifications etc., along with sufficiently accurate positional
    information.''

In this paper we present a list of \rrls, with $V_{\rm max}\leq 12.5$ mag, 
all present in the GCVS, but now  with absolutely reliable positions. 
It includes 286 Hipparcos and Tycho stars (ESA 1997; \hip), of which 104
only in Tyc2, as well as  273 in the GSC (Lasker \ea (1996)), 
and a few USNO-A2 stars 
(the USNO-A2 is published by Monet \ea (1998)).
Our list contains the names used in each of these catalogues. 

\section{Comparing catalogues}
\label{comparing}

As  a first step we began with an identification of RR\,Lyr stars in the GCVS 
with a brightness at maximum of $ V_{\rm max} \leq$ 12.5 mag.
We then proceeded to inspect their positions by looking up 
the original finding charts. 
It emerged from that already that several \rrls\ were not identified properly. 
Thus a full inspection of all the basic data was required. 

Because it was impossible to observe all \rrls\ ourselves, 
and especially stars from the southern hemisphere could not be observed by us,
we made a literature search to find the missing ones.
We followed very strict principles. 
First, papers which give a finding chart or image identifying the variable
by observing its light-curve were regarded as a reliable reference.
Second, we used papers giving the Tyc2 or GSC number of the star, 
by the same criteria. 
Beside this we tried to find the original record of discovery
and the finding chart given there. 
It takes considerable effort to do so, 
especially if only  very old finding notes exist, 
which are not available electronically.

 All images, finding charts and records of the discoverer were examined
in the same way as our own observations. We compared them
with the DSS and overlaid them  with Tyc2 and GSC from the 
VizieR catalogue service using the Aladin interactive sky atlas.
However, for some GCVS stars the identification was still elusive 
and it was decided to go and observe the fields anew to 
properly identify the particular \rrls.

\section{Observation and Identification of RR Lyrae Stars}
\label{observation}

The best way to identify \rrls\ in the field of stars is  
to observe the fields and look for the star with \rrl\  like variability.

\begin{figure}[t]
\centering
\epsfig{file=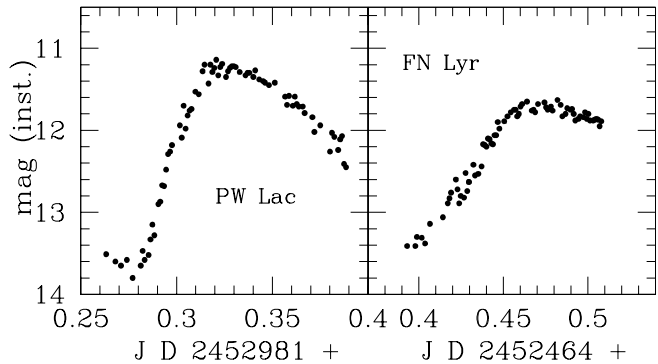,scale=0.45}
\caption[]{Light-curve of PW Lac (left panel) and FN Lyr (right panel) 
observed at the  observatory of G. Maintz. 
Such data was used to identify the RR Lyrae stars in their fields.}
\label{FNLyrPWLac.fig}
\end{figure}

For 34 fields time series  were obtained in Bonn, 
using a $8''$ Schmidt-cassegrain Meade telescope 
 and an OES Alpha-Mini CCD-camera with an IRR-UV-Cut filter (B\&W 486). 
This equipment is fully sufficient for taking  series of images for
several hours allowing to define the light-curve.
Usually about every minute an image was taken. For an example of
light-curves see Fig.\,\ref{FNLyrPWLac.fig}.

The data  reduction was 
made with the OES-Fleischmann software provided with the camera. 
 We identified the \rrl\, easily by its characteristic variability.
We then compared our image with the DSS as described in 
Sect.\,\ref{comparing}.
With this procedure we extracted name and coordinates of the star 
in question from the Tyc2 and GSC catalogue, respectively.

\begin{figure}[t]
\centering
\epsfig{figure=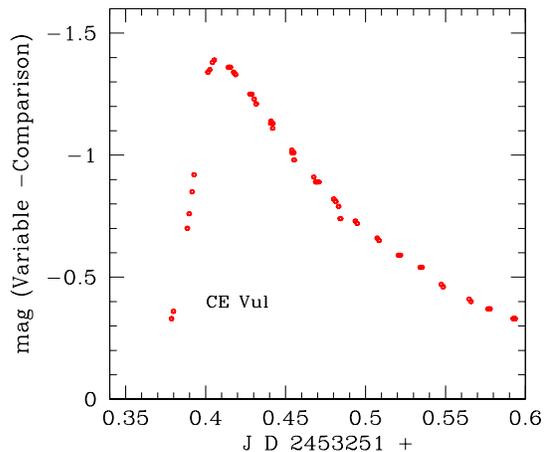,height=6cm,width=7cm}
\caption[]
{Light-curve of CE Vul observed at the observatory ``Hoher List'' 
of the Sternwarte of the Univ.\ of Bonn. 
Such data was required to identify RR Lyrae stars in their fields.}
\label{CEVul.fig}
\end{figure}

For 22 stars time series  were observed at the Ob\-ser\-va\-tory Ho\-her 
List with the 1\,m Cassegrain-Telescope using the Cassegrain focus with 
focal reducer (f=368 cm) and the HoLiCam 2048${\times}$2048 pix CCD camera. 
For the reduction of these CCD-images we used standard CCD reduction
routines  (DAOPHOT under IRAF) to determine positions and
 magnitudes of all stars on the frames.
The \rrl\ in question was identified by its light-curve 
(for an example see Fig.\,\ref{CEVul.fig}) 
and its identification number and coordinates were obtained by  
comparing with the DSS overlaying with the Tyc2 or GSC, too.

\section{The final catalogue}

We give all \rrls\ with $V_{\rm max}\leq 12.5$ mag
(according to GCVS)  in one list, but we mention 
that the 186  \rrls\  from the \hip\ which  are 
included  in our catalogue,
are taken from \hip\ directly needing no further identification.

Our final catalogue contains 577 \rrls\ with the following information 
in the columns
\begin{description}
\itemsep=0pt\parsep=0pt\parskip=0pt\labelsep=0pt
\item 1: Name of the variable (GCVS).
\item 2+3: Catalogue (H=\hip, T=Tyc2, G=GSC, UA2 = USNO-A2) + number; 
an asterisk points at an individual note to the star.\footnote{The source of identification is available from gmaintz@astro.uni-bonn.de} 
\item 4+5: $\alpha,\delta$ (2000) given in the particular catalogue. 
\item 6: $V$ magnitude in  maximum light. 
\item 7: class of \rrl\, based on light-curve. 
\item 8+9: first epoch and period.
\item 10: date of last observed maximum light known by us. 
\item 11: source of first epoch and period if not taken from GCVS  or \hip. 
\end{description}

If the star is taken from \hip,  the magnitude in  maximum light, type of 
variability, first epoch and period are from the \hip\ Variability Annex. 
If this information comes from another catalogue, it is from the GCVS, 
except when new information for  first epoch and period 
are available in the literature. 

The date of the last observation of maximum light 
mostly comes from amateur observers.
They regularly observe variables of all kinds and they obtain light-curves of 
\rrls\ as well, achieving correct maximum epochs.
This column is, although not complete, 
very  useful to make predictions of maximum light.

If a star could not be identified
in its field,  because no second source for identification 
besides the GCVS was available (and no variable was
found in our observation of the field) our catalogue 
has 999999999 as number. 
This happened in eight cases.
The position given for these stars is taken from the GCVS.
If there are no data for period or first epoch in the GCVS
and no information is found otherwise,  its first epoch and
period is given as 9999999.9999 and  9.99999999, respectively.

For 6 stars without Tyc2 or GSC number we give identifications and 
coordinates (2000) from the USNO-A2.0. 
Magnitudes and elements are taken from the GCVS.

 For three stars (DM And, CQ Lac and UZ Eri) being members of Tyc2, there 
are  no proper motions available in the Tyc2. 
Therefore those stars are not useful for further 
investigations based on Tyc2.

X\,CMi is identified as GSC 168-406 while  
Schmidt \ea  (1995) identified it as GSC 168-562. 
However, comparing their finding chart with the chart from   
Tsesevisch \& Kazanasmas (1971), 
 we found that in both charts there is only one star at the  
position of X CMi while the POSS gives a close pair. 
GSC\,168-562 is the preceding star of the pair  
and is very faint (15.25 mag according to GSC III aJ). 
That is why we adopt that X\,CMi is the following star (168-406) 
with a magnitude of 12.34 mag (GSC III aJ).

 Mennessier \&  Colome (2002; hereafter MC) gave results 
for 172 RR Lyrae stars identified to be Tyc2 stars.
They did the identification blindly, 
using  the GCVS as source of the positions and magnitudes of the \rrls.
They were successful in finding many \rrls\ at Tyc2 star positions.
But they did not find all and, as it turned out during our search,
their list is wanting since they misidentified a large fraction of their stars.

For 78 stars of their 172 we agree with the identifications of MC. 
In 93 cases we found their stars being misidentified. 
We found 27 Tyc2 stars which are not given by MC.
  From the misidentified ones we found 84 in the GSC and 3 in the USNO-A2,  
while for 3 Tyc2 stars
 (NO\,Cas,  V672\,Aql, V1823\,Cyg\footnote{According to  IBVS Nr 4997, 
V1823 Cyg is  a W-UMa star   and not a RR Lyrae star. 
It is of course not included in our catalogue.}) 
we found another identification number than given by MC.
The misidentification mostly happens, when  the star was fainter than
the magnitudes given in the GCVS. 
 Of the MC paper stars, 3 are not mentioned in our list:
V1823\,Cyg (see IBVS 4997), 
UU\,Cam (Schmidt \ea 1995), CE\,Aqr (A.Paschke priv com.),
 {\sl because they are not} \rrls. 
14 stars of their list are fainter than our limiting magnitude,
but we added these stars to our list for completeness.

\section{Summary}
We present a catalogue listing  
561 \rrls\ with $V_{\rm max}  \leq 12.5$ mag (plus 16 fainter ones)
with exact coordinates.  
Most of them come from the GCVS. 
21 have recently been recovered ( of the new found \rrls\ 2
have a  $V_{\rm max}$ 
of 12.52 and 12.53 mag respectively).
27 \rrls\, are previously unknown as   members of Tyc2.  
Data of first epoch and period are given if available.
 The dates of the last maxima observed come mostly from amateur observers and 
(even if this data is incomplete) allow to calculate trust-able predictions.

As an example, 
the beginning of the catalogue is given at the end of this paper.\footnote{The complete Table 1 is  available in electronic form at CDS via anonymous ftp to cdsarc.u-strasbg.fr (130.79.128.5) or via http://cdsweb.u-strasbg.fr/cgi-bin/qcat?/A+A/}

\nopagebreak

\acknowledgements
This research has made use of the Simbad database, operated at CDS, 
Strasbourg, France. 
We thank Karl W\"alke and Wolfgang Grimm who donated 
the Atlas of Finding Charts for Variable Stars by  Tsesevisch and Kazanasmas. 
We thank Franz Agerer, Wolfgang Moschner, Konstantin von Poschinger 
and K. Bernhard for sending several identifications of stars from 
their observations.
We thank Anton Paschke for sending maximum light times of recent observations.

\begin{sidewaystable*}
\caption{561 RR-Lyrae Stars with maximum light $\le 12.5 $ mag according to GCVS plus 16 fainter ones.}
\label{RRauszug.tab}
\setlength{\tabcolsep}{1.15mm}
\begin{tabular}{lcrrrrlllll}
\hline
Name              &  Cat &  Nr         & $\alpha_{2000}$ &   $\delta_{2000}$ & $ V_{\rm max}$ &    type    & First Ep.  & Per. & last observ.  & source   \\
of star            &      &                &                &                   &  (mag)         &            &  (JD)        & (d) & (JD)            & of Period   \\
 \hline
SW   And      & H    &      1878 &  00 23 43.090 &  29 24 03.62 &  9.22 & RRab & 2448500.0384 & 0.4422620000 & 53260.1300 &     \\
XX   And      & H    &      6029 &  01 17 27.415 &  38 57 02.03 & 10.20 & RRab & 2448500.6580 & 0.7227550000 & 53045.3410 &     \\
AT   And      & H    &    116958 &  23 42 30.832 &  43 00 51.66 & 10.51 & RRab & 2448500.2310 & 0.6169170000 & 53232.5870 &     \\
BK   And      & G    & 323500127 &  23 35 06.022 &  41 06 10.51 & 12.50 & RRab & 2429146.4500 & 0.4215985000 & 51321.8076 &     \\
CI   And      & H    &      8939 &  01 55 08.294 &  43 45 56.47 & 11.78 & RRab & 2448500.0022 & 0.4847280000 & 52695.3060 &     \\
DK   And      & G*   & 364500701 &  23 28 45.910 &  50 34 29.35 & 12.50 & RR:  & 2429130.4070 & 0.2436553000 &            &     \\
DM   And      & T    & 277300253 &  23 32 00.707 &  35 11 48.87 & 12.40 & RRab & 2435717.4310 & 0.6303890000 & 51335.8300 &     \\
DR   And      & T    & 228600352 &  01 05 10.707 &  34 13 06.24 & 12.00 & RRab & 2437220.3190 & 0.5631180000 & 52620.4385 &     \\
DU   And      & G    & 283600362 &  02 30 31.337 &  40 50 33.11 & 12.50 & RR   & 2436051.4500 & 0.6067160000 & 51469.4641 &     \\
OV   And      & T    & 278701874 &  00 20 44.206 &  40 49 40.80 & 10.40 & RRab & 2439026.4780 & 0.4705810000 & 52983.4265 & 1   \\
V395 And      & H    &    117111 &  23 44 32.143 &  46 22 48.59 &  7.57 & RRc: & 2448500.2660 & 0.3423280000 &            &     \\
WY   Ant      & H    &     50289 &  10 16 04.946 & $-$29 43 42.41 & 10.40 & RRab & 2448500.5339 & 0.5743410000 & 51869.6630 &     \\
SY   Aps      & G*   & 926500960 &  14 39 24.130 & $-$72 49 35.62 & 12.20 & RR:  & 2451904.0650 & 0.2789100000 &            &  2  \\
TY   Aps      & H    &     72444 &  14 48 50.012 & $-$71 19 41.88 & 11.27 & RRab & 2448500.4217 & 0.5016950000 & 52823.5250 &     \\
TY   Aps      & H    &     72444 &  14 48 50.012 & $-$71 19 41.88 & 11.27 & RRab & 2448500.4217 & 0.5016950000 & 52823.5250 &     \\
UY   Aps   & UA2 & 0150 14253525 &  14 59 34.440 & $-$71 47 53.77 & 12.00 & RRab & 2425326.5400 & 0.4825200000 & 51921.9800 &     \\
VX   Aps      & G    & 942900076 &  15 59 56.690 & $-$75 13 21.00 & 11.50 & RRab & 2434239.3610 & 0.4845780000 & 51927.5010 &     \\
 \dots        &   \dots &  \dots &         \dots &          \dots & \dots &\dots &        \dots &        \dots &      \dots &     \\
 \dots        &   \dots &  \dots &         \dots &          \dots & \dots &\dots &        \dots &        \dots &      \dots &     \\
 \noalign{\smallskip}
 \hline
 \noalign{\smallskip}
\end{tabular}
\vspace{5mm}
\begin{tabbing}
  1  \quad \= Rossiger, S. \& Busch, H., 1988, Mitt. Verand. Sterne 11, 133   \= \kill
        \> The sources of first epoch and  period if not from \hip\ or GCVS: \\
    1   \> Rossiger, S. \& Busch, H., 1988, Mitt. Verand. Sterne 11, 133   \\
    2   \> A. Paschke priv. communication \\
  \dots                           \>         \dots                   \\
  \dots                           \>         \dots                   \\
\end {tabbing}
\begin{tabbing}
      SY Aps   \quad \=  maybe  Ecl. Var. EW (Pasc  \= \kill
Individual notes: \>  \\
      DK And       \>  type EW is also possible.         \>  A. Paschke priv. communication \\
      SY Aps       \>  type EW is also possible.          \> A. Paschke priv. communication \\
        \dots      \>     \dots                           \>         \dots                   \\
        \dots      \>     \dots                           \>         \dots                   \\
\end{tabbing}
Online-Data are available under http://www.astro.uni-bonn.de/$\sim$gmaintz

\end{sidewaystable*}


\begin{thebibliography}{}
\bibitem[]{}Altmann, M., \& de Boer, K.S. 2000,  A\&A 353, 135 
\bibitem[]{}Altmann, M., Edelmann, H., \&  de Boer, K.S. 2004,  A\&A 414, 181 
\bibitem[]{}Beers, T.C., Chiba, M., Joshii, Y., \ea   2000, AJ 119, 2688  
\bibitem[]{}de Boer, K. S., Aguilar Sanchez, Y., Altmann, M., \ea  1997, A\&A 327, 577  
\bibitem[]{}Dambis, A. K., \& Rastorguev, A.S. 2001, AstL 27, 108 
\bibitem[]{} ESA, 1997, The Hipparcos and Tycho Catalogues, ESA SP\-1200 and its Variability Annex: Periodic variables  
\bibitem[]{}Fernley, J., Barnes, T.G., Skillen, I. \ea  1998, A\&A 330, 515  
\bibitem[]{}H{\o}g, E., Fabricius, C., Makarov, V.V.,  \ea    2000, Tycho-2 catalogue  (Tyc2)    
\bibitem[]{}Kaempf, K.S., de Boer, K.S, \& Altmann, M. 2005  A\&A 432, 879
\bibitem[]{}Kholopov, P.N., Samus, N.N., Durlevich, O.V., \ea  1990, General Catalogue of Variable Stars, 4rd ed., vol.IV, Nauka, Moscow. (GCVS)           
\bibitem[]{}Lasker, B.M., Russel, J.N., \& Jenkner, H. 1996, The HST Guide Star Catalogue, Version 1.2 
\bibitem[]{}Layden, A.C. 1994, AJ 108, 1016 
\bibitem[]{}Layden, A.C., Hanson, R.B., Hawley, S.L., \ea 1996, AJ 112, 2110    
\bibitem[]{}Maintz, G., de Boer, K.S.  2005, A\&A, submitted
\bibitem[]{}Martin, J.C.,  \& Morrison, H.L. 1998, AJ 116, 1724  
\bibitem[]{}Mennessier, M.O., \& Colomé, J.  2002, A\&A  390,  173  
\bibitem[]{}Schmidt, E.G., Chab, J.R., \& Reiswig, D.E. 1995, AJ 109, 1239 
\bibitem[]{}  Monet, D., Bird, A., Canzian, B.,  \ea 1998, A Catalogue of Astrometric Standards (USNO-A V2.0)
\bibitem[]{}Tsesevisch, V.P., \& Kazanasmas, M.S. 1971, Atlas of Finding Charts for Variable Stars, Moscow  

\end{thebibliography}
\end{document}